# $_nC_k$ sequences and their difference sequences

Asbjørn Brændeland

***Abstract***

A $_nC_k$ sequence is a sequence of *n*-bit numbers with *k* bits set. Given such a sequence *C*, the difference sequence *D* of *C* is subject to certain regularities that make it possible to generate *D* in 2|*C*| time, and, hence, to generate *C* in 3|*C*| time. What induces the regularities of *D* appear clearly in the bit representation of *C*.

Every subset of a set *S* can be represented by a set of *indices*, which can in turn be represented by *bit-fields* where the positions of the bits that are set, are indices of the elements of *S*. Let $n = |S|$. Then $2^n - 1$ represent *S*, and $P = \{0, 1, \ldots, 2^n - 1\}$ represent the powerset of *S*.

Now we want the set $C \subset P$ such that every number in *C* has bit count *k*. *C* can be produced by simply selecting these numbers from *P*, but if we only want *C*, this is too expensive. The cardinality of *C* is $\binom{n}{k}$, and even for $k = \lfloor n/2 \rfloor$, which gives the largest *C*, |*C*| is only a fraction of |*P*|, a fraction that gets smaller as *n* grows, and if *k* is close to 1 or *n*, the fraction is very small. E.g. for $n = 50$ we have $2^n - 1 = 1125899906842623$, $\binom{n}{25} = 126410606437752$, $\binom{n}{4} = \binom{n}{46} = 230300$ and $\binom{n}{2} = \binom{n}{48} = 1225$.

Another way to generate *C* is to take the partial sums of its difference sequence *D*. Due to certain regularities, *D* can be generated in 2|*C*| time. What induces these regularities, becomes apparent when we look at the bit representation of *C*, as demonstrated in Figure 1.

To avoid confusion we let *bar*, rather than *one*, denote *a bit that is set*. An $_nC_k$ *number* is an *n*-bit field with *k* bars and an $_nC_k$ *sequence* is a sequence of $_nC_k$ numbers.

A $_nC_k$ sequence can be partitioned into sub sequences according to the position of the leftmost bar $b_L$. The length of a sub sequence is determined by the number of positions and bars to the right of $b_L$. E.g. in the $_7C_3$ sub sequence 19, 21, 22, 25, 26, 28, where $b_L$ is in zero based position 4, the two remaining bars fill 4 positions in 6 different ways. The $_5C_1$ sequence has 5 sub sequences of length 1, the $_6C_2$ sequence has 5 sub sequences of lengths 1, 2, 3, 4 and 5, and the $_7C_3$ sequence has 5 sub sequences of lengths 1, 3, 6, 10 and 15. Notice that (1, 3, 6, 10, 15) are the partial sums of (1, 2, 3, 4, 5), which are the partial sums of (1, 1, 1, 1, 1).

Let $C_{n,k}$ be an $_nC_k$ sequence and let $D_{n,k}$ be the corresponding difference sequence. Then the sub sequence of length *s* in $D_{n,k}$, corresponding to the sub sequences of length *s* in $C_{n,k}$, contains the first *s* numbers in $D_{n-1,k-1}$, with $2^{k-2}$ added to the last number. With $b_L$ staying in the same position, only the remaining $k - 1$ bars contribute to the differences—hence the repetitions. The added $2^{k-2}$ is required to get to the start of the next sub sequence of $C_{n,k}$, where $b_L$ is moved one position to the left, and the other $k - 1$ bars are moved to the beginning, i.e. the right side, of the bit field.

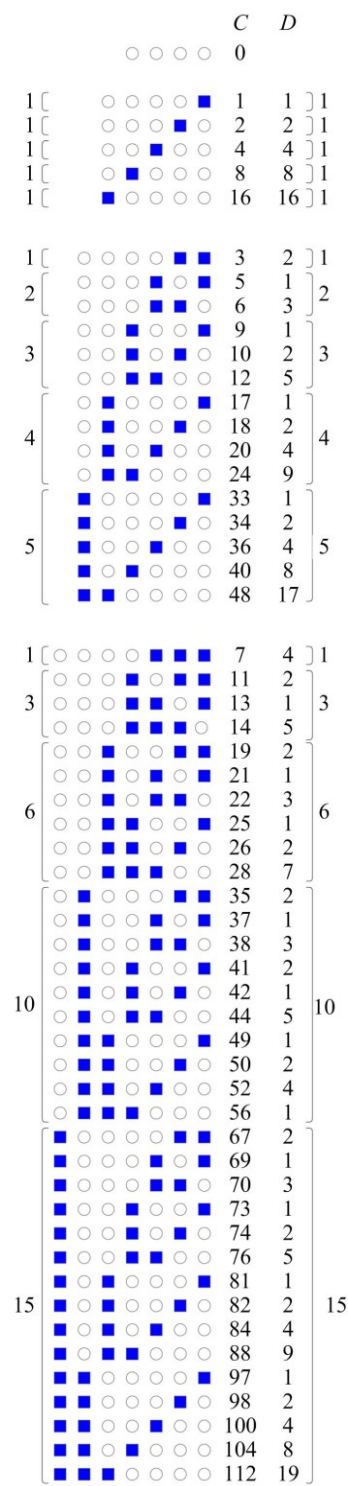

Figure 1.

With regard to an implementation, we can look at these regularities as follows:

The sequences below contain the 5-bits, 6-bits and 7-bits numbers with 1, 2 and 3 bits set, respectively.

    1, 2, 4, 8, 16. (1a)

    3, 5, 6, 9, 10, 12, 17, 18, 20, 24, 33, 34, 36, 40, 48. (2a)

    7, 11, 13, 14, 19, 21, 22, 25, 26, 28, 35, 37, 38, 41, 42, 44, 49, 50, (3a)
    52, 56, 67, 69, 70, 73, 74, 76, 81, 82, 84, 88, 97, 98, 100, 104, 112.

The corresponding difference sequences are

    1, 2, 4, 8. (1b)

    2, 1, 3, 1, 2, 5, 1, 2, 4, 9, 1, 2, 4, 8. (2b)

    4, 2, 1, 5, 2, 1, 3, 1, 2, 7, 2, 1, 3, 1, 2, 5, 1, 2, 4, 11, 2, 1, 3, 1, 2, 5, 1, 2, 4, 9, 1, 2, 4, 8. (3b)

Notice that (1b) is repeated at the end of (2b) and (2b) is repeated at the end of (3b).

We now split (2b) and (3b), up to the repeated parts, into four segments each, shown here preceded by their segment lengths

    1    2,                                      1    4,
    2    1, 3,                                  3    2, 1, 5,
    3    1, 2, 5,            (2c)        6    2, 1, 3, 1, 2, 7,        (3c)
    4    1, 2, 4, 9,                      10    2, 1, 3, 1, 2, 5, 1, 2, 4, 11.

In (2c) each part is a segment of (1b) with $2^0 = 1$ added to the last term, and in (3c), each part is a segment of (2b) with $2^1 = 2$ added to the last term. Notice that the length column in (3c) contains the partial sums of the length column in (2c). Notice also that the second numbers in the length columns in (2c) and (3c) give the respective bit counts (number of bars) of (2a) and (3a). The power to which 2 is raised in the addend of the last number in each segment, equals the bit count − 2.

Inspired by this, we make a new length column containing the partial sums of the length column in (3c) and a corresponding list of segments of (3b), adding $2^2 = 4$ to the last term in each segment.

    1    8,
    4    4, 2, 1, 9,
    10    4, 2, 1, 5, 2, 1, 3, 1, 2, 11, (4b)
    20    4, 2, 1, 5, 2, 1, 3, 1, 2, 7, 2, 1, 3, 1, 2, 5, 1, 2, 4, 15.

If we then add (3b), it is a fair assumption that the result would be the difference sequence of the 8-bit numbers with 4 bits set, and, lo and behold, when we prepend 15, i.e., the first number with 4 bits set, and take the partial sums of the result (the complement of the difference sequence, as it were), we get

    15, 23, 27, 29, 30, 39, 43, 45, 46, 51, 53, 54, 57, 58, 60, 71, 75, 77, 78, 83, 85, 86, 89, 90, 92, 99, 101, 102,
    105, 106, 108, 113, 114, 116, 120, 135, 139, 141, 142, 147, 149, 150, 153, 154, 156, 163, 165, 166, 169, (4a)
    170, 172, 177, 178, 180, 184, 195, 197, 198, 201, 202, 204, 209, 210, 212, 216, 225, 226, 228, 232, 240,

which is exactly the 8-bit numbers with 4 bits set.

The width of the initial bit field, in this case 4 (see top of Figure 1), determines the width of the last bit field, so if we for instance had wanted the 12-bit numbers with 4 bits set, we would have had to start with 8 bits—and, generally, to get the $n$-bit numbers with $k$ bits set, we start with $n − k$ bits.

The following *list*-based *Racket*-implementation illustrates the principle, but the repeated use of **append** makes it a bit slower than necessary. (Racket is a dialect of Scheme, which is a dialect of Lisp.)

```
(define (powers-of-two n) (map (lambda (p) (expt 2 p)) (enumerate 0 n)))   ; †

(define (partial-sums sequence)
  (define (iterate seq sums)
    (if (null? seq)
        (reverse sums)                          ; sums was built in reverse order by means of cons, to save time.
        (iterate (rest seq) (cons (+ (first seq) (first sums)) sums))))
  (if (or (null? seq) (null? (rest seq)))
      seq
      (iterate (rest seq) (list (first seq)))))

(define (n-choose-k n k)
  (define (join-segments segment-lengths prev-diff-sequence)
    (define addend (expt 2 (- (second segment-lengths) 2)))
    (append (flatten (map (lambda (lng)          ; Join the segments with the augmented last number.
                            (append (take prev-diff-sequence (- lng 1))
                                    (list (+ (list-ref prev-diff-sequence (- lng 1))
                                             addend))))
                          segment-lengths))
            prev-diff-sequence))                 ; Finally, append the entire previous difference sequence.
  (define (iterate segment-lengths difference-sequence)
    (if (> (second segment-lengths) k)           ; The partial sums mechanism makes the
                                                 ; second segment length usable as a counter.
        difference-sequence
        (iterate (partial-sums segment-lengths)
                 (join-segments segment-lengths difference-sequence))))
  (define initial-seg-lengths (make-list (- n k) 1))   ; (1 1 ... 1), i.e., $n - k$ ones.
  (define choose-1 (powers-of-two (- n k 1)))          ; (1 2 4 ... $2^{n-(k+1)}$)
  (partial-sums (cons (- (expt 2 k) 1)                 ; Prepend the first number with $k$ bits set.
                      (iterate (partial-sums initial-seg-lengths)
                               choose-1))))
```

A *vector*-based implementation is more optimal (allthough a little more cumbersome to write in Racket). E.g., for $n = 50$ and $k = 7$ the vector-based implementation is about 50 times faster than the list-based one.

```
(define (vector-partial-sums! v) ; ‡
  (define len (vector-length v))
  (define (iterate! i)
    (and (< i len)
         (vector-set! v i (+ (vector-ref v (- i 1)) (vector-ref v i)))
         (iterate! (+ i 1))))
  (and (> len 1) (iterate! 1))
  v)
```

† `enumerate` is not a Racket procedure. It takes two arguments *a* and *b* and returns the sequence *a*, *a* + 1, …, *b*.
‡ It is common practice in Lisp to let the name of a destructive procedure end with "!". This has nothing to do with the factorial.

```
(define (vector-sum v)
  (define (iterate i sum) (if (< i 0) sum (iterate (- i 1) (+ sum (vector-ref v i)))))
  (iterate (- (vector-length v) 1) 0))

(define (vector-n-choose-k n k)
  (define (copy-segments! lengths prev-seq)
    (define (copy-segment! vector-1 offset-1 vector-2 offset-2 len)
      (define (iterate! i)
        (and (< i len)
             (vector-set! vector-2 (+ offset-2 i) (vector-ref vector-1 (+ offset-1 i)))
             (iterate! (+ i 1))))
      (iterate! 0)
      vector-2)
    (define n-lengths (vector-length lengths))
    (define new-segs-length (vector-sum lengths))
    (define prev-sequence-length (vector-length prev-seq))
    (define new-sequence (make-vector (+ new-segs-length prev-sequence-length)))
    (define addend (expt 2 (- (vector-ref lengths 1) 2)))
    (define (iterate! i offset-2)
      (and (< i n-lengths)
           (let ((lng (vector-ref lengths i)))
             (copy-segment! prev-seq 1 new-sequence offset-2 (- lng 1))
             (vector-set! new-sequence
                          (+ offset-2 (- lng 1))
                          (+ (vector-ref prev-seq lng) addend))
             (iterate! (+ i 1) (+ offset-2 lng)))))
    (iterate! 0 1)
    (copy-segment! prev-seq 1 new-sequence (+ new-segs-length 1) (- prev-sequence-length 1)))
  (define (calc-diffs i lengths prev-seq)
    (if (< i (- k 1))
        (let ((lengths (vector-partial-sums! lengths)))
          (calc-diffs (+ i 1) lengths (copy-segments! lengths prev-seq)))
        prev-seq))
  (define initial-seg-lengths (make-vector (- n k) 1))
  (define choose-1 (list->vector (cons 0                                  ; Make room for the first number with k bits set.
                                       (powers-of-two (- n k 1)))))
  (define diffs (calc-diffs 0 initial-seg-lengths choose-1)) ; Generate entire difference sequence.
  (vector-set! diffs 0 (- (expt 2 k) 1))                     ; Place the first number with k bits set.
  (vector-partial-sums! diffs))                              ; Return the choose_{n,k} sequence.
```

Of course, a vector-based implementation would have been even more efficient in a vector oriented language such as C. Notice, however, that Racket does not limit the size of numbers—which is a prerequisite for having *n-choose-k* handling large *n*.